\documentclass[aps,prx,showpacs,twocolumn,amsmath,amssymb]{revtex4}

\usepackage[colorlinks,citecolor=blue,linkcolor=Fuchsia]{hyperref}
\usepackage[usenames,dvipsnames,svgnames,table]{xcolor}

\usepackage{graphicx}
\usepackage{Jonasmacros}
\usepackage{bm}
\usepackage{mathptmx}
\usepackage{txfonts}

\begin{document}
\date{\today}
\title{Inelastic impurity scattering induced spin texture and topological transitions in surface electron waves}

\author{J. Fransson}
\email{Jonas.Fransson@physics.uu.se}
\affiliation{Department of Physics and Astronomy, Uppsala University, Box 516, SE-751 21\ \ Uppsala}



\begin{abstract}
Inelastic scattering off magnetic impurities in a spin chiral two-dimensional electron gas, e.g., Rashba system, is shown to generate topological changes in the spin texture of the electron waves emanating from the scattering center. While elastic scattering gives rise to a purely in-plane spin texture for an in-plane magnetic scattering potential, out-of-plane components emerge upon activation of inelastic scattering processes. This property leads to a possibility to make controlled transitions between trivial and non-trivial topology of the spin texture.
\end{abstract}
\maketitle

\section{Introduction}
Controlling the mechanisms behind magnetic interactions provides general tools for designing specific properties which, in turn, can be utilized to probe fundamental physical phenomena. While the properties of both ferromagnetism and anti-ferromagnetism can be described in terms of, e.g., Heisenberg and Ising models \cite{heisenberg1926,dirac1926,heisenberg1928,ising1925}, the introduction of spin-orbit (SO) interaction opens possibilities to generate spin textures with non-trivial topology, e.g., spin spirals and helical configurations as well as merons and skyrmions \cite{bogdanov2001,yu2010,heinze2011,pereiro2014}.

While mainly considered as ground state properties, magnetic properties in excited states may be different and it is therefore interesting, both fundamentally and technologically, to be in control of the transitions between various configurations. Such control is efficiently provided by inelastic scattering processes. Measurements of inelastic transitions open up a route to investigate the excitation spectrum of physical systems. There has been a growing activity in elucidating inelastic scattering processes in quantum systems using various experimental techniques. An incomplete list includes inelastic neutron \cite{prassides1991,christianson2008} and X-ray \cite{lee2005,saiz2008} scattering, transport through break junctions \cite{park2000,wang2004,rau2014}, and scanning tunneling microscopy (STM) with spin-polarized (SP-STM) \cite{meier2008,zhou2010}, non spin-polarized tip \cite{stipe1998,hirjibehedin2006,hirjibehedin2007,otte2008,otte2009,balashov2009,khajetoorians2011,khajetoorians2013,donati2013}, or superconducting STM \cite{heinrich2013}.

Surface imaging of scattering states can be performed, e.g., by using STM to probe the spatial spectral density variations at a given energy. Friedel oscillations caused by elastic scattering processes emerge around defects adsorbed onto a surface \cite{hasegawa1993,sprunger1997}. Recent studies also predict {\em inelastic Friedel oscillations} emerging from vibrating \cite{balatsky2003,fransson2007,she2013,fransson2013} and magnetic \cite{balatsky2003,fransson2012} impurities, of which the former was detected for dimers of meta-dichlorobenzene on Au surface \cite{gawronski2010}.

\begin{figure}[t]
\begin{center}
\includegraphics[width=.99\columnwidth]{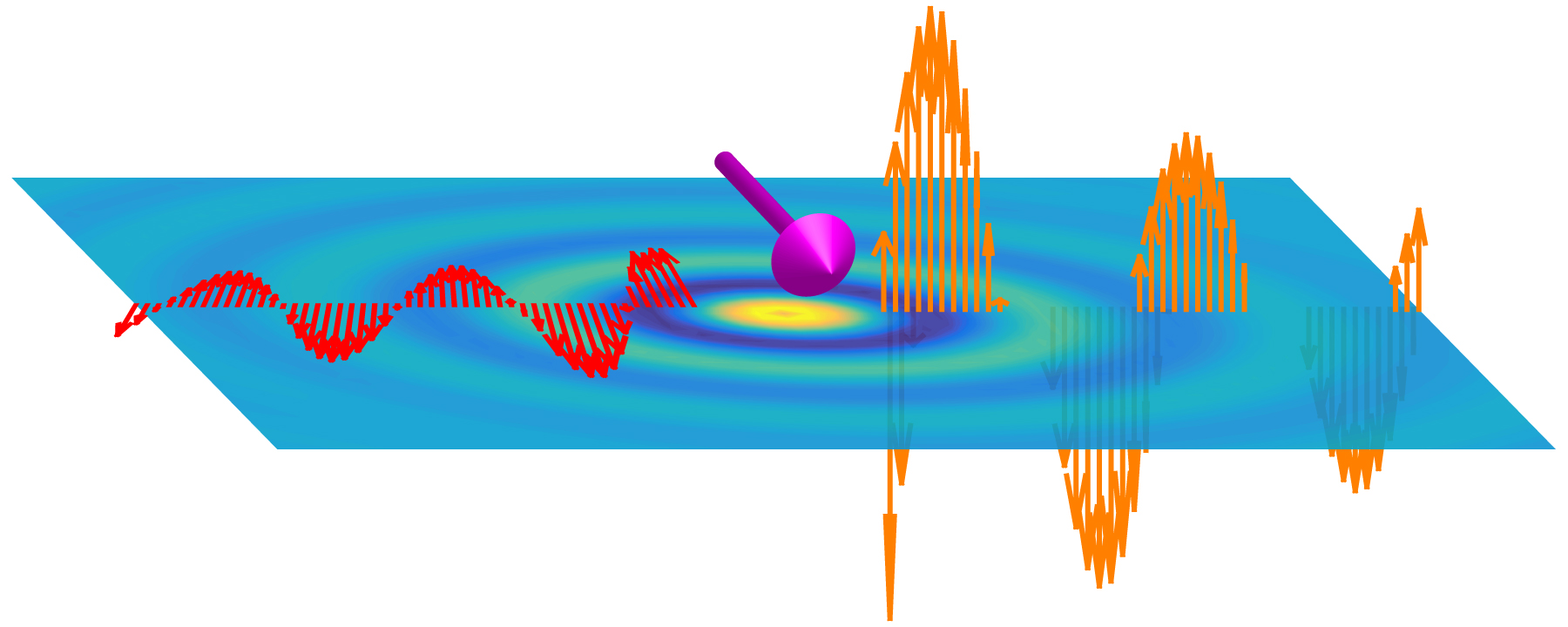}
\end{center}
\caption{Illustration of the elastic (left) and inelastic (right) Friedel oscillations emerging from the in-plane magnetic scattering center. 
}
\label{fig-HW}
\end{figure}

In this paper, we study the spin texture emerging around a magnetic defect embedded in a two-dimensional spin chiral electronic structure, something which can be realized by, e.g., magnetic adatom on a metallic Rashba surface. In particular, we consider the influence of inelastic scattering on the spin texture superimposed on the Friedel oscillations emanating from the scattering center. It is demonstrated that the spin texture caused by elastic and inelastic scattering may have dramatically different properties. While elastic scattering off an in-plane magnetic adsorbant induces a sole in-plane spin texture in the electronic structure, see Fig. \ref{fig-HW}, inelastic scattering comprises out-of-plane components, Fig. \ref{fig-HW}. The inelastic scattering can be stimulated only by adding or removing an energy quantum that corresponds to the inelastic transition energy of the impurity spin, which opens for controlled transitioning between the two topologically distinct textures, making controlled experimental, e.g., STM, studies of different topologies under essentially similar conditions possible.

Furthermore, controlled tuning between different topological states is potentially important for quantum information technology since the physical properties corresponding to the different topological states are essentially distinct. The capability to deliberately switch between different topologies by turning on and off spin-inelastic scattering by means of external force, e.g., voltage bias, open for all-electrical switches and sensor applications.

The paper is organized as follows. In Sec. \ref{sec-model} we introduce a model system for which the presented study is based on. In Sec. \ref{sec-elastic} we work through some details of the elastic scattering properties while the inelastic scattering properties are discussed in Sec. \ref{sec-inelastic}. The paper is summarized and concluded in Sec. \ref{sec-summary}.

\section{Model}
\label{sec-model}
As an example of the predictions, we consider a general spin $\bfS(t)=\bfS(\bfr_0,t)$ located at the position $\bfr_0$ on a spin chiral metallic surface. In terms of the spinor $\Psi_\bfk=(\cs{\bfk\up}\ \cs{\bfk\down})^T$, the surface electrons can be modeled by $\Hamil_\text{surf}=\sum_\bfk\Psi_\bfk^\dagger[\dote{\bfk}\sigma^0+\alpha(\bfk\times\hat{\bf z})\cdot\bfsigma]\Psi_\bfk$, where $\dote{\bfk}$ denotes the single electron energy at the momentum $\bfk$ and $\alpha$ defines the Rashba spin-orbit coupling. Here, $\sigma^0$ is the $2\times2$ identity matrix and $\bfsigma$ is the vector of Pauli matrices. The interaction between the localized magnetic moment and the surface states is captured within the Kondo model $\Hamil_K=J_K\bfs(\bfr_0,t)\cdot\bfS(t)$, where $J_K$ is the effective Kondo exchange parameter. The electron spin is denoted by $\bfs(\bfr,t)=\Psi^\dagger(\bfr,t)\bfsigma\Psi(\bfr,t)$ with $\Psi(\bfr,t)=\sum_\bfk\Psi_\bfk(t)e^{i\bfk\cdot\bfr}$. Considering the Rashba model is no severe restriction since the results can be understood also in view of the linear Dresselhaus model, c.f., Refs.  \onlinecite{koralek2009,studer2010}.

In our calculated examples below, the localized spin moment is described by the Hamiltonian $\Hamil_S=DS_z^2+E[S_+^2+S_-^2]/2$, where the anisotropy fields $D$ and $E$ account for the effective interaction between the localized spin moment and the surface electrons. In terms of this model, we define the $2S+1$ eigenstates $\ket{\alpha}$ and corresponding eigenenergies $E_\alpha$. We then obtain the expectation value $\av{\bfS}=\av{S_z}\hat{\bf z}$ along its local spin quantization axis which, however, may be non-collinear with the spin quantization axis defined by the surface electrons. We write $\av{\bfS}=\av{S_z}\hat{\bf n}$ to indicate the direction of the local spin moment within the spin quantization axis of the surface electrons.

We connect our results for the local density of electron states (DOS) to the (differential) conductance of the STM measurements through the formula $dI(\bfr,V)/dV\propto n_0N(\bfr,eV)+\bfm_0\cdot\bfM(\bfr,eV)$, where $n_0$ ($N(\bfr,eV)$) and $\bfm_0$ ($\bfM(\bfr,eV)$) is the DOS and spin-polarization of the tip (substrate) \cite{tersoff1983,wortmann2001,fransson2010}.

We calculate the real space Green function (GF) for the surface electrons $\bfG(\bfr,\bfr';z)=\int\bfG(\bfk,\bfk';z)e^{i\bfk\cdot\bfr-i\bfk'\cdot\bfr'}d\bfk d\bfk'/(2\pi)^4$, where $\bfG(\bfk,\bfk';z)$ is a $2\times2$ matrix in spin space. In general, we can write $\bfG=G_0\sigma^0+\bfG_1\cdot\bfsigma$, where $\bfG_0$ and $\bfG_1$ represent the charge and magnetic components, respectively. In this representation $N(\bfr,eV)\equiv-\im\, {\rm sp}\, \bfG(\bfr,\bfr;eV)/\pi=-2\im G_0(\bfr,\bfr;eV)/\pi$ and $\bfM(\bfr,eV)\equiv-\im\, {\rm sp}\, \bfsigma\bfG(\bfr,\bfr;eV)/2\pi=-\im\bfG_1(\bfr,\bfr;eV)/\pi$, where ${\rm sp}$ is the trace over spin 1/2 space.

\section{Elastic scattering}
\label{sec-elastic}
First we construct a {\it bare} GF $\bfG^{(0)}$ which contains the spin-polarization induced by the localized magnetic moment. Within the $T$-matrix formulation \cite{fiete2001,franssonNL2010,fransson2014}, we obtain the bare GF which includes scattering off the defect to all orders in the scattering potential $\bfV=V_0\sigma_0+\bfsigma\cdot\bfDelta(z)$ \cite{black-schaffer2015} comprising the spin-independent  and spin-dependent contributions $V_0$ and $\bfDelta(z)=v_uJ_K\av{\bfS}(z)$, respectively. We write
\begin{subequations}
\begin{align}
\bfG^{(0)}(\bfk,\bfk')=&
	\delta(\bfk-\bfk')\bfg(\bfk)
	+e^{-i\bfk\cdot\bfr_0}\bfg(\bfk)\bfT\bfg(\bfk')e^{i\bfk'\cdot\bfr_0},
\\
\bfT=&
	\Bigl(\bfV^{-1}-\bfg(\bfr=0)\Bigr)^{-1}
	=
	(T_0\sigma^0+\bfsigma\cdot\bfDelta)/t_0,
\\
T_0=&
	V_0+i(V_0^2-|\bfDelta|^2)N_0/2,
\\
t_0=&
	1-(V_0^2-|\bfDelta|^2)(N_0/2)^2+iV_0N_0,
\end{align}
\end{subequations}
where ($k=|\bfk|$)
\begin{align}
\bfg(\bfk,z)=&
	\frac{(z-\dote{\bfk})\sigma^0+\alpha[\bfk\times\hat{\bf z}]\cdot\bfsigma}{(z-\dote{\bfk})^2-\alpha^2k^2},
\end{align}
is the GF for the free surface states, giving $\bfg(\bfr=0)=\int\bfg(\bfk)d\bfk/(2\pi)^2=-iN_0\sigma_0/2$, $N_0=m/\hbar^2$. The corresponding retarded real space GF is given by $\bfg^r(\bfr,\omega)=g_0(\bfr,\omega)\sigma^0+\bfg_1(\bfr,\omega)\cdot\bfsigma$, where $g_0(\bfr,\omega)=-iN_0\sum_{s=\pm}\kappa_sH_0^{(1)}(\kappa_sr)/4\kappa$ and $\bfg_1(\bfr,\omega)=-N_0\sum_{s=\pm}s\kappa_sH_1^{(1)}(\kappa_sr)\hat\bfr\times\hat{\bf z}/4\kappa$, with $\kappa_\pm=\kappa\pm\alpha N_0$, $\kappa=\sqrt{2N_0(\omega+\alpha^2N_0/2)}$, $\omega$ is the energy relative to the Fermi level, $r=|\bfr|$, and $\hat{\bfr}=\bfr/|\bfr|$, and where $H_n^{(1)}(z)$ is the Hankel function. This is in agreement with \cite{lounis2012}, which also means that our results below can be understood as superpositions of two wave like components with the different wave vectors $\kappa_\pm$. The magnetic structure of the free surface states can be written $\bfm(\bfr,\omega)=-\im\bfg_1(\bfr,\omega)/\pi=(\bfm_\perp(\bfr,\omega),m_z(\bfr,\omega)$) with vanishing spin-polarization $m_z(\bfr,\omega)=0$. There is, nonetheless, an in-plane magnetic structure $\bfm_\perp(\bfr,\omega)\neq0$ with $\nabla\times\bfm_\perp(\bfr,\omega)\neq0$ which manifests the chirality. The topology of the magnetic structure is trivial, however, since the topological number $q=\int\hat\bfm\cdot(\partial_x\hat\bfm\times\partial_y\hat\bfm)d\bfr/4\pi=0$.

The correction $\delta\bfG^{(0)}(\bfk,\bfk')=e^{-i\bfk\cdot\bfr_0}\bfg(\bfk)\bfT\bfg(\bfk')e^{i\bfk'\cdot\bfr_0}$ is partitioned into charge and magnetic contributions $\delta\bfG^{(0)}(\bfk,\bfk')=\delta g_0(\bfk,\bfk')\sigma^0+\delta\bfg_1(\bfk,\bfk')\cdot\bfsigma$, where
\begin{subequations}
\label{eq-deltag}
\begin{align}
\delta g_0(\bfk,\bfk')=&
	a(\bfk)
	g_0(\bfk')
	+
	{\bf b}(\bfk)
	\cdot
	\bfg_1(\bfk'),
\label{eq-deltag0}
\\
\delta\bfg_1(\bfk,\bfk')=&
	a(\bfk)
	\bfg_1(\bfk')
	+
	{\bf b}(\bfk)
	g_0(\bfk')
	+
	i
	{\bf b}(\bfk)
	\times
	\bfg_1(\bfk')
	,
\label{eq-deltag1}
\end{align}
\end{subequations}
using the notation $a(\bfk)=[g_0(\bfk)T_0+\bfg_1(\bfk)\cdot\bfDelta]/t_0$ and ${\bf b}(\bfk)=[g_0(\bfk)\bfDelta+\bfg_1(\bfk)T_0+i\bfg_1(\bfk)\times\bfDelta]/t_0$. The interplay between the intrinsic magnetic texture of the surface electrons and the elastic impurity scattering leads to that an induced non-trivial spin texture emerges around the adsorbant. First, in absence of the spin-dependent scattering potential, $J_K\rightarrow0$, the magnetic correction $\delta\bfg_1(\bfk,\bfk')=T_0[g_0(\bfk)\bfg_1(\bfk')+\bfg_1(\bfk)g_0(\bfk')+i\bfg_1(\bfk)\times\bfg_1(\bfk')]/t_0$, showing the absence of back-scattering, $\delta\bfg_1(\bfk,-\bfk)=0$, while skew-scattering is allowed, $\delta\bfg_1(\bfk,\bfk')\neq0$, $\bfk'\neq-\bfk$, as expected \cite{fransson2014}. Only by inclusion of the spin-dependent potential, back scattering becomes allowed.

The overall magnetic structure is nevertheless modified significantly by the presence of the impurity potential, which becomes very evident in the real space description. Define $\bfG^{(0)}(\bfr,\bfr')\equiv\bfg(\bfr-\bfr')+\delta\bfG^{(0)}(\bfr,\bfr')=G^{(0)}_0(\bfr,\bfr')\sigma^0+\bfG^{(0)}_1(\bfr,\bfr')\cdot\bfsigma$, where the correction term $\delta\bfG^{(0)}(\bfr,\bfr')=\bfg(\bfr-\bfr_0)\bfT\bfg(\bfr_0-\bfr')=\delta g_0(\bfr,\bfr')\sigma^0+\delta\bfg_1(\bfr,\bfr')\cdot\bfsigma$. The components $\delta g_0(\bfr,\bfr')$ and $\delta \bfg_1(\bfr,\bfr')$ are given by Eq. (\ref{eq-deltag}) by replacing $\bfk\rightarrow\bfR_0=\bfr-\bfr_0$ and $\bfk'\rightarrow-\bfR_0'=\bfr_0-\bfr'$. We find the correction to the DOS $\delta N(\bfr,\omega)$ and magnetic structure $\delta\bfM(\bfr,\omega)$ as
\begin{subequations}
\label{eq-elastid}
\begin{align}
\delta N^{(0)}(\bfr,\omega)=&
	-\im
	\frac{2}{\pi t_0}
	\biggl(
		T_0
		\Bigl(
			g_0^2(\bfR_0)
			+g_1^2(\bfR_0)
		\Bigr)
\nonumber\\&
		+2g_0(\bfR_0)\bfg_1(\bfR_0)\cdot\bfDelta
	\biggr)
\label{eq-elasticdN}
\\
\delta\bfM^{(0)}(\bfr,\omega)=&
	-\im
	\frac{2}{\pi t_0}
	\left(
		\frac{
			g_0^2(\bfR_0)
			-g_1^2(\bfR_0)
			}
			{2}
		\bfDelta
		+
		T_0g_0(\bfR_0)\bfg_1(\bfR_0)
	\right.
\nonumber\\&
	\left.
	\vphantom{\frac{g_0^2(\bfR_0)-g_1^2(\bfR_0)}{2}}
		+
		\bfg_1(\bfR_0)\cdot\bfDelta\, \bfg_1(\bfR_0)
	\right).
\label{eq-elasticdM}
\end{align}
\end{subequations}

The scattering generates Friedel oscillations emerging from the impurity site, which is a result of the broken translation symmetry caused by the impurity. The standing wave pattern is a result of the oscillatory nature of both $g_0(\bfr)$ and $\bfg_1(\bfr)$, see above. It is also clear that the Friedel oscillations arise from both spin-independent and -dependent scattering potentials.

\begin{figure}[t]
\begin{center}
\includegraphics[width=.99\columnwidth]{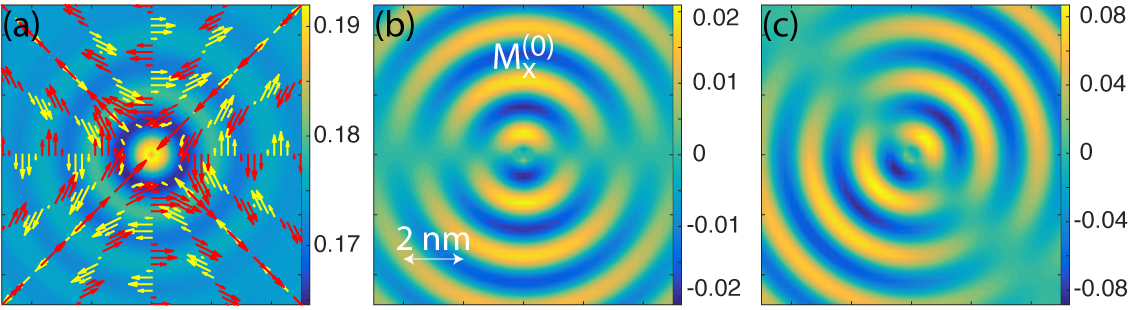}
\end{center}
\caption{Calculated (a) $nN(\bfr,V)$, (b) $M^{(0)}_x(\bfr)$, and (c) $\bfm_0\cdot\bfM^{(0)}(\bfr,V)$, around an adsorbed spin moment. In (a), the arrows indicate the direction of the spin texture. Here, we have used the Rashba SO coupling $\alpha_0=0.012/$\AA\ \cite{ast2007}, the effective electron mass $m^*=0.26m_e$, with free electron mass $m_e$, temperature $T=0.1$ K, quadratic energy dispersion for the surface states $\dote{\bfk}-E_F=k^2/2N_0+E_0$, with $E_0\simeq-0.45$ eV. The tip is assumed to have a magnetic moment $\bfm_\tip=3n_0(1,1,\sqrt{2})/8$, where $n_0$ is the tip DOS at the Fermi level.}
\label{fig-Fig1}
\end{figure}

Concerning the magnetic density, we first consider a non-magnetic impurity ($\bfDelta=0$) for which we have $\delta\bfM^{(0)}(\bfr,\omega)=-2\im T_0g_0(\bfR_0)\bfg_1(\bfR_0)/\pi t_0$. This expression shows that the spin texture emanating from the scattering center retains its in-plane character around the adsorbant, see Fig. \ref{fig-Fig1} (a), which shows the computed conductance around the impurity. The spin texture (arrows) is inherited in the Friedel oscillations emerging from the scattering center. However, due to the oscillatory properties of both the charge ($g_0$) and spin ($\bfg_1$) densities and as function of the distance $|\bfR_0|$, the spin texture changes direction from clockwise to counter clockwise quasi-periodically with the Friedel oscillations, which is illustrated by $M^{(0)}_x(\bfr,V)$ in Fig. \ref{fig-Fig1} (b), $M^{(0)}_y(\bfr,V)$ is equal to $M^{(0)}_x(\bfr,V)$ but rotated clockwise $90^\circ$. In addition to the alternating directions of the spin texture, it also continuously alternates between being purely radial (along $y=\pm x$) and purely azimuthal (along $y=0$ and $x=0$).
The spin texture will be visible only in an SP-STM measurement where the tip magnetic moment $\bfm_0$ comprises an in-plane component, see Fig. \ref{fig-Fig1} (c), which shows the component $\bfm_0\cdot\bfM^{(0)}(\bfr,V)$ to the conductance. Despite the non-trivial spin texture emanating from the scattering center the corresponding magnetic topology is nevertheless trivial since again $Q^{(0)}=\int\hat\bfM^{(0)}\cdot[\partial_x\hat\bfM^{(0)}\times\partial_y\hat\bfM^{(0)}]d\bfr/4\pi=0$.

\begin{figure}[t]
\begin{center}
\includegraphics[width=.99\columnwidth]{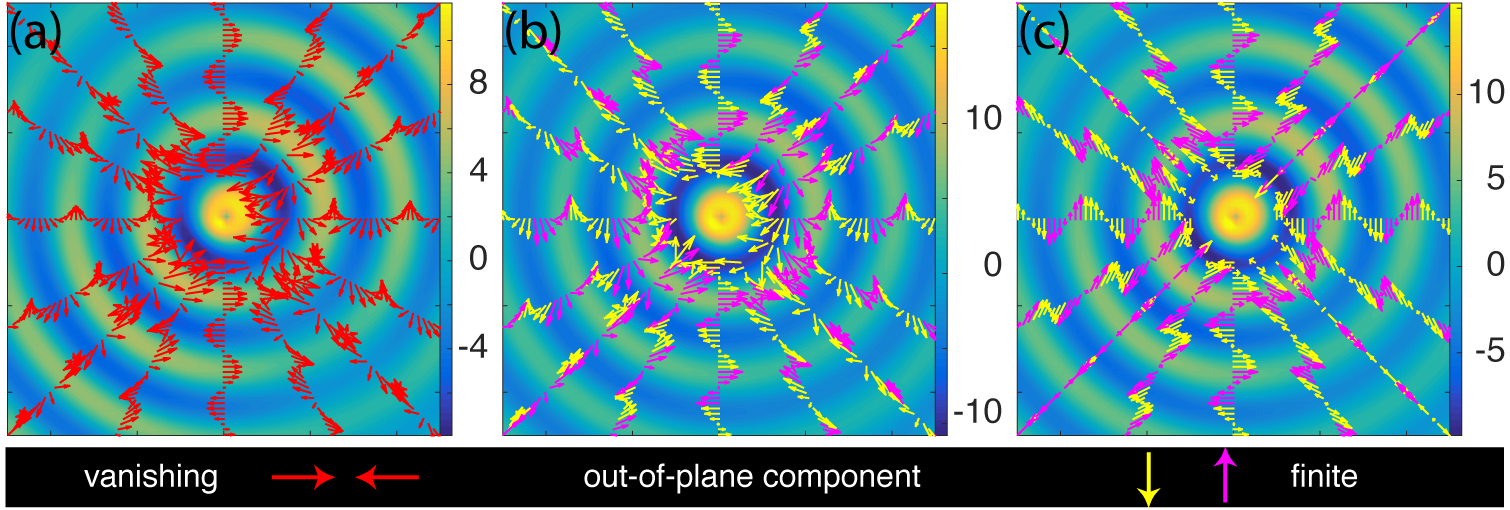}
\end{center}
\caption{In-plane projection of the calculated spin texture (arrow colors indicate vanishing -- red, negative -- yellow, and positive -- magenta, $z$-component), overlayered on top of $\bfm_0\cdot\bfM^{(0)}(\bfr,V)$, emanating from the magnetic scattering center. Here, $|\bfDelta|/V_0=1/2$ with (a) $\hat{\av{\bfS}}=(1,0,0)$, (b) $\hat{\av{\bfS}}=(1,0,1)/\sqrt{2}$, and (c)  $\hat{\av{\bfS}}=(0,0,1)$. (d) Topological number $Q^{(0)}$ as function of the polar angle defined between the $\av{\bfS}$ and the spin quantization axis of the surface electrons. Other parameters are as in Fig. \ref{fig-Fig1}.}
\label{fig-Fig2}
\end{figure}

The picture changes dramatically in presence of a spin-dependent scattering potential, $\bfDelta\neq0$. First, if $\bfDelta\perp\hat{\bf z}$, only the in-plane spin texture is modified by the spin-dependent impurity scattering while the out-of-plane component remains vanishing, see Fig. \ref{fig-Fig2} (a), which illustrates the spin texture of the surface electrons for a local spin $\hat{\av{\bfS}}=(1,0,0)$. Accordingly the topological number $Q^{(0)}=0$, as can be seen in the left end of Fig. \ref{fig-Fig2} (d). Close to the scattering center, the spin texture tends to be aligned with the local magnetic moment while its direction is again alternating between clockwise and counter clockwise with the Friedel oscillations away from the scattering center. The in-plane orientation of the local spin moment breaks the rotation symmetry of the induced spin texture and prevents the formation of lines with purely radial texture.

Inclusion of an out-of-plane component to the scattering center, see Fig. \ref{fig-Fig2} (b, c) where $\hat{\av{\bfS}}=(1,0,1)/\sqrt{2}$ and $(0,0,1)$, respectively, generates a finite out-of-plane component in $\delta\bfM^{(0)}(\bfr,V)$, in addition to the in-plane spin texture. Hence, the resulting spin texture emerging from the scattering center is in general a spin spiral, or a spin helix. For the latter spin moment, $\hat{\av{\bfS}}=(0,0,1)$, regions of purely radial spin texture form along $y=\pm x$, since the rotation symmetry is preserved in the system. The topology of the induced magnetic structure is now also non-trivial, see Fig. \ref{fig-Fig2} (d),\cite{TN} due to the presence of the finite out-of-plane component which continuously alternates between the negative and positive $z$-direction. We verify the skyrmion-like surface electron waves found in Ref. \onlinecite{lounis2012}. We notice, however, that these previous results only account for the radial component of the wave and, hence, incorrectly describe the full spin texture which actually emanates helically from the scattering center.


%

\section{Inelastic scattering}
\label{sec-inelastic}

\begin{figure}[b]
\begin{center}
\includegraphics[width=.99\columnwidth]{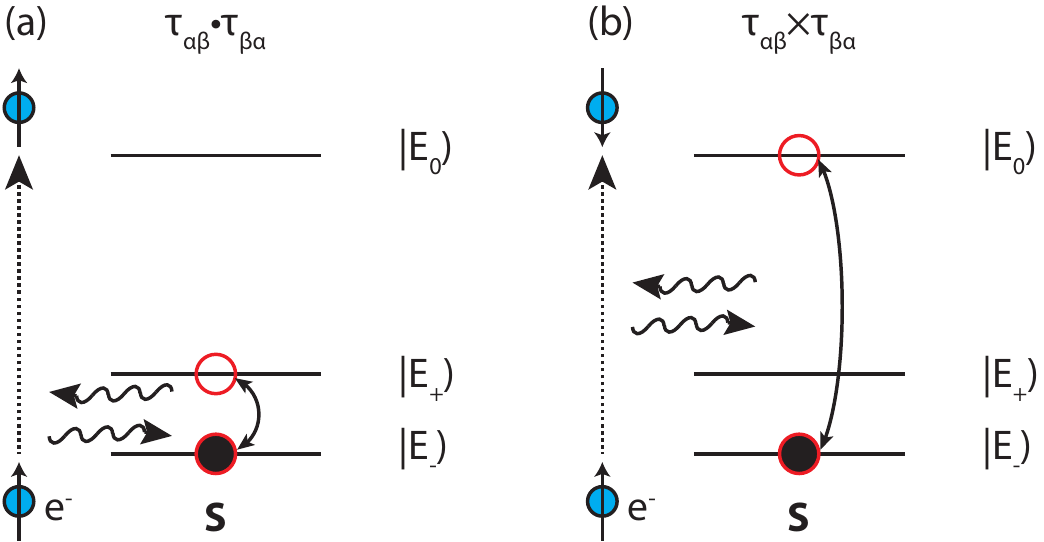}
\end{center}
\caption{Simple illustration of the spin-inelastic transitions for an $S=1$ system with $D<0$ and $E\neq0$. The tunneling electron, $e^-$, interacts and exchange energy and angular momentum with the local spin moment $\bfS$. Processes involve (a) energy exchange while electron spin is conserved and (b) energy and spin exchange accompanied by electron spin flip.}
\label{fig-transitions}
\end{figure}

Next, we consider effects of spin-inelastic scattering, something which can be stimulated by, e.g., the voltage bias in STM measurements. In second order perturbation theory with respect to $J_K$, effects from spin-inelastic scattering off the localized spin moments influences the surface state. We define the self-energy
\begin{align}
\Sigma_{\sigma\sigma'}(\bfr_0;z)=&
	-\frac{J_K^2}{\beta}
	\sum_\nu\bfsigma_{\sigma s}\cdot\chi(z_\nu)\cdot\bfsigma_{s'\sigma'}
	G_{ss'}^{(0)}(\bfr_0;z-z_\nu)
\label{eq-exp} 
\end{align}
($\beta^{-1}=k_BT$) where $\bfG^{(0)}(\bfr;z)\equiv\bfG^{(0)}(\bfr,\bfr;z)$, whereas the spin-spin propagator $\chi(z)$ is the Fourier transform of $\chi(t,t')=\eqgr{\bfS(t)}{\bfS(t')}$, which describes the correlations of the local spin moment at different instances in time.
In real space, we make the expansion
\begin{align}
\bfG(\bfr,\bfr')\approx&
	\bfG^{(0)}(\bfr,\bfr')
			+
			\bfG^{(0)}(\bfr,\bfr_0)\bfSigma(\bfr_0)\bfG^{(0)}(\bfr_0,\bfr').
\label{eq-dressedGF}
\end{align}

As we focus on the main effects from spin-inelastic scattering, we calculate the spin-spin propagator in the atomic limit $\chi(t,t')\approx(-i){\rm sp}\bftau\calG(t,t')\bftau\calG(t',t)$, where $\calG$ is the single electron matrix GF defined in basis $\{\ket{\alpha}\}$, whereas $\bftau$ defines the vector of corresponding spin matrices. Assuming that the basis $\{\ket{\alpha}\}$ is an eigenbasis, we can write the GF $\calG(z)=\{\delta_{\alpha\beta}/(z-E_\alpha)\}_{\alpha\beta}$.
Then, the correction term $\delta\bfG=\bfG^{(0)}\bfSigma\bfG^{(0)}$ acquires the form
\begin{subequations}
\begin{align}
\delta\bfG(\bfr,\bfr')=&
			\bfG^{(0)}(\bfr,\bfr_0)
			\int
				\bfA^{(0)}(\dote{})
				\calL(\dote{})
			\frac{d\dote{}}{\pi}
			\bfG^{(0)}(\bfr_0,\bfr'),
\\
\calL(\dote{})=&
	J_K^2
	\sum_{\alpha\beta}
		(\bftau_{\alpha\beta}\cdot\bftau_{\beta\alpha}\sigma^0+i[\bftau_{\alpha\beta}\times\bftau_{\beta\alpha}]\cdot\bfsigma)
\nonumber\\&\times
			\frac{f(E_\alpha)f(-E_\beta)-[f(E_\alpha)-f(E_\beta)]f(\dote{})}
			{z-E_\alpha+E_\beta+\dote{}},
\end{align}
\end{subequations}
where $\bfA^{(0)}(\dote{})=\im\bfG^{(0),r}(\bfr_0;\dote{})=-N_0(\sigma^0+N_0\im{\bfT}/2)/2$, and $f(x)$ is the Fermi function.

\begin{figure}[b]
\begin{center}
\includegraphics[width=.99\columnwidth]{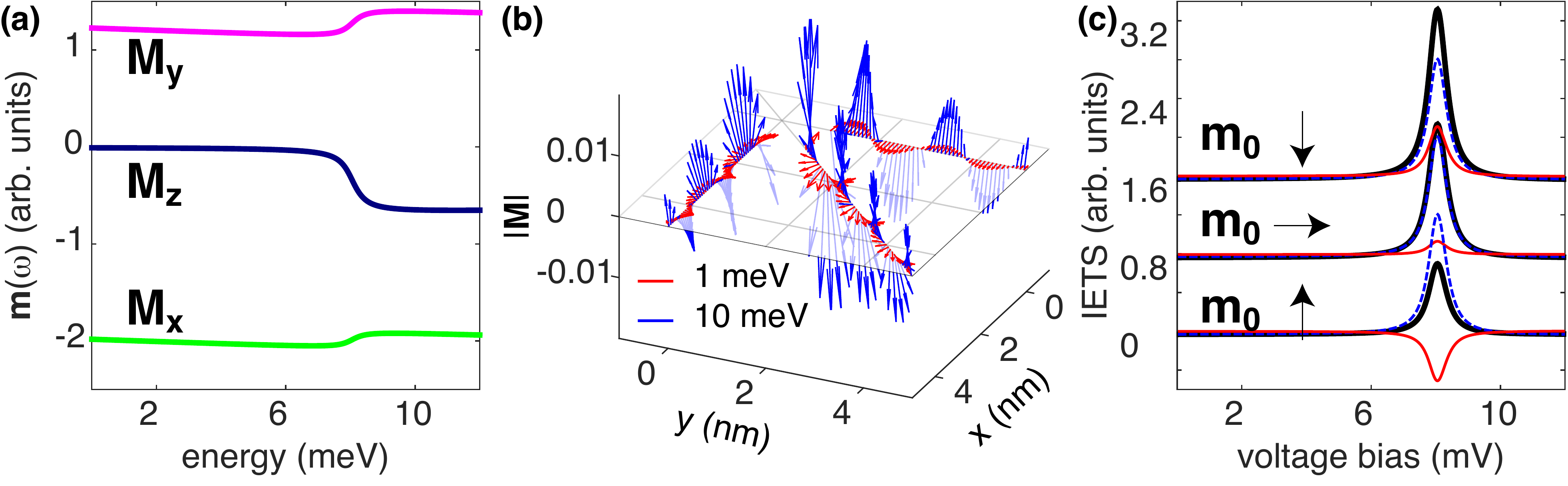}
\end{center}
\caption{Calculated (a) magnetic densities $M_i(\bfr,\omega)$, $i=x,y,z$ at $\bfr=1\hat{\bf x}$ nm (bold) and topological number $Q$ (dashed), (b) the induced spin texture along different spatial directions, as function of energy, and (c) $d^2I(\bfr,V)/dV^2$ (bold) and its corresponding non-magnetic (dashed) and magnetic (faint) contributions. Here, $S=1$ using $D=-20$ meV, $E=|D|/5$. Other parameters as in Fig. \ref{fig-Fig1}.}
\label{fig-Fig3}
\end{figure}

The factor $\calL$ indicates that the correction term $\delta\bfG$ is active for energies near and larger than the transition energy $E_\alpha-E_\beta$, such that processes corresponding to both emission and absorption of the energy quanta $E_\alpha-E_\beta$ are accounted for. This energy is exchanged with the surrounding electronic density. Moreover, processes that are either spin conserving or spin non-conserving of the local spin moment are accounted for by the products $\bftau_{\alpha\beta}\cdot\bftau_{\beta\alpha}$ and $\bftau_{\alpha\beta}\times\bftau_{\beta\alpha}$, respectively. The processes involved here and described by the correction $\delta\bfG$ are illustrated in Fig. \ref{fig-transitions}. The former processes are stimulated by electrons that either emit or absorb the appropriate energy quantum to the spin system without exchanging any angular momentum and, hence, stimulate transitions between spin states for which the difference in angular momentum $\Delta m_z=0$. The latter processes correspond to transitions between states which differ in angular momentum by $\Delta m_z=\pm1$. These are stimulated by electrons which energy transfer is accompanied by a spin flip.

\begin{figure}[b]
\begin{center}
\includegraphics[width=.99\columnwidth]{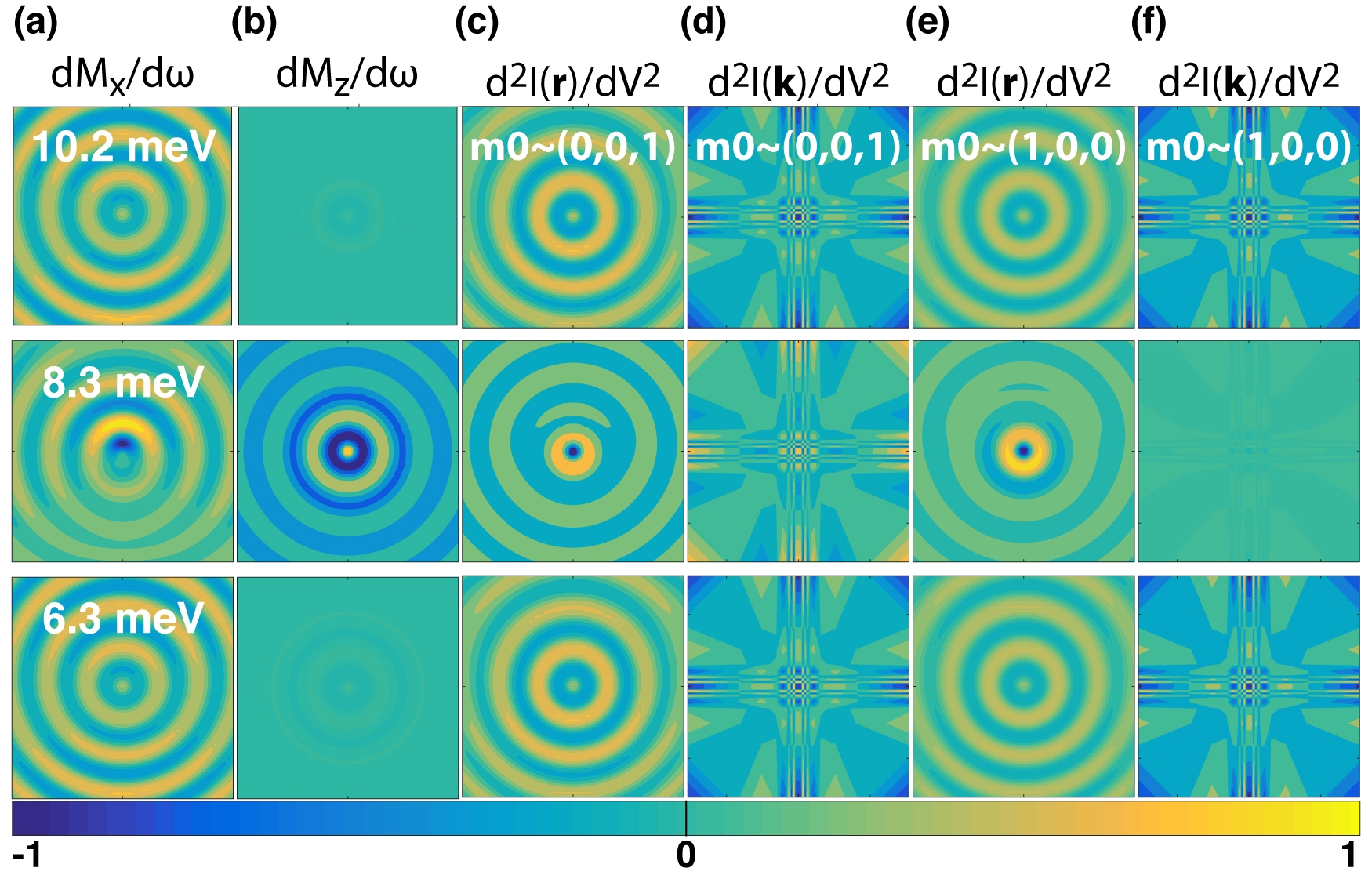}
\end{center}
\caption{Differential magnetic densities (a) $\partial_\omega M_x(\bfr,\omega)$ and (b) $\partial_\omega M_z(\bfr,\omega)$, (c), (d) spatial and $k$-space IETS maps for tip magnetic moment $\bfm_0\sim(0,0,1)$ and (e), (f) IETS maps for $\bfm_0\sim(1,0,0)$, for the energies $6.3,\ 8.3,\ 10.2$ meV. Other parameters as in Fig. \ref{fig-Fig3}.}
\label{fig-Fig4}
\end{figure}

The terminology \emph{spin-inelastic scattering} is reasonable since in these processes the localized spin moment undergoes transitions (a) between states with different energy and (b) between states with different energy and angular momentum, see Fig. \ref{fig-transitions}. In the former case, also referred to as spin conserving, for a spin $S=1$ system the states $\ket{E\pm}=\alpha_\pm\ket{S=1,m_z=-1}+\beta_\pm\ket{S=1,m_z=1}$ which, hence, open for transitions between these states as long as the energy difference between are fed into/taken out of the local spin moment. This energy is exchanged with the surrounding electrons, in particular for, e.g., STM this energy can be exchanged with the electrons in the tunneling current. In the latter case, also referred to as spin non-conserving, the local spin moment undergoes transitions between states not only with different energy but also with different angular momentum, hence, which is exchanged with the surrounding electrons under spin-flip processes such that the difference in angular momentum of $\Delta m_z=\pm1$ in the localized spin moment is compensated. The processes can be generalized to other spin systems.

In this context it is particularly interesting to study the induced spin texture, represented by the correction $\delta\bfM(\bfr,\omega)$, see the Appendix, emerging from the scattering center at varying energies. Among the expected properties for inelastic scattering in the present context are peak/dips in the inelastic electron tunneling spectroscopy (IETS), $d^2I(V)/dV^2$, and strong voltage bias variations in the inelastic imaging contrast, $d^2I(\bfr)/dV^2$. However, at the spin-inelastic transition energies $E_\alpha-E_\beta$, the orientation of the spin texture may undergo dramatic changes. Most conspicuously, an in-plane scattering center generates a strong out-of-plane spin texture at the inelastic transition energy. This can be seen from the presence of a contribution of the type
\begin{align}
\re
	\int
		\biggl\{
			\bfG^{(0),r}_1(\bfr,\bfr_0)\times[\im\bfG^{(0),r}_1(\bfr_0;\dote{})]
			\Bigl(
				\calL^r_0(\dote{})G^{(0),r}_0(\bfr_0,\bfr)
\nonumber\\
				+
				\calL^r_1(\dote{})\cdot\bfG^{(0),r}_1(\bfr_0,\bfr)
			\Bigr)
		\biggr\}
	d\dote{},
\label{eq-dMoop}
\end{align}
in the dressed magnetic structure $\delta\bfM(\bfr,\omega)$. As we have seen above for elastic scattering, the spin texture described by $\bfG^{(0)}_1$ is purely in-plane for an in-plane magnetic scattering center. Then, the cross product in the contribution in Eq. (\ref{eq-dMoop}) generates a magnetic component which is solely out-of-plane, however, it is only effective for energies that are sufficiently large to support inelastic transitions. The plots in Fig. \ref{fig-Fig3} (a) illustrate the energy dependence of the magnetic vector $\bfM$ at $\bfr=1\hat{\bf x}$ nm, for an $S=1$ impurity with $D=-20$ meV and $E=|D|/5$. For low enough energies, the out-of-plane component of the spin texture is vanishing, or negligible, while the in-plane components are finite. Likewise, the topological number $Q=\int\hat\bfM\cdot[\partial_x\hat\bfM\times\partial_y\hat\bfM]d\bfr/4\pi=0$, which is essentially obtained numerically as well, see Fig. \ref{fig-Fig3} (a) (dashed). At the energy for the spin-inelastic transition, here $\sim8$ meV, all components show a step due to the activation of inelastic processes. In particular, the out-of-plane component becomes finite at this energy, something which indicates a structural change in the induced spin texture and is confirmed by the rapid change in the topological number. This change is depicted in Fig. \ref{fig-Fig3} (b), where the low energy spin texture is completely in-plane (red) while there is a helical type of spatial modulation for energies sufficiently large to assist spin-inelastic scattering (blue). The plots in Fig. \ref{fig-Fig3} (c) show the total $d^2I/dV^2$ (bold) and its non-magnetic (dashed) and magnetic (faint) components for tip magnetic moments $\bfm_0$ pointing in-plane and out-of-plane. The asymmetry between the different signatures can be used to determine the orientation of the local spin texture.

The spatial variations of the spin texture can, in principle, be resolved in terms of IETS mapping as function of the applied voltage bias. The plots in Fig. \ref{fig-Fig4} (a,b) display the differential magnetic densities $\partial_\omega M_x(\bfr,\omega)$ and $\partial_\omega M_z(\bfr,\omega)$, respectively, for three energies around the inelastic transition. Specifically, these sequences show a dramatic variation in the out-of-plane component close to the inelastic transition, while the in-plane component is much less influenced. This clearly illustrates that the out-of-plane magnetism undergoes a strong change upon activation of the inelastic scattering. The rather short range of the induced texture, which is a result of the strong (weak) out-of-plane (in-plane) variations, is transferred over to the $d^2I(\bfr,V)/dV^2$, Fig. \ref{fig-Fig4} (c, e), such that the transport signature becomes focused to the scattering center.
This property can be efficiently analyzed in the Fourier transform $d^2I(\bfk,V)/dV^2$, see Fig. \ref{fig-Fig4} (d, f), since focusing the spatial signature to a small area leads to a weak signature in momentum space. By using a spin-polarized tip with $\bfm_0$ out-of-plane, however, it can noticed that the momentum space IETS map tends to become inverted near the inelastic transition energy, see Fig. \ref{fig-Fig4} (d), which indicates a strong variation in the out-of-plane component of the spin texture.
This is in contrast to the case when $\bfm_0$ is in-plane, Fig. \ref{fig-Fig4} (f), where the signature is completely absent at the inelastic transition energy. This can be understood since the variations in the out-of-plane spin texture, which dominates the overall signature at the inelastic transition energy, has a different symmetry than the spin-polarized tip and cannot be picked up in the measurement.

Fourier transform spectroscopy using STM has been successfully used previously to acquire spectroscopical information that is not accessible directly in the spatial data, e.g., \onlinecite{sprunger1997,hoffman2002,zhu2004,strozecka2011}. Hence, the theoretically predicted inelastic scattering induced transition between a trivial and non-trivial topology of the spin texture emerging around the magnetic adatom, is therefore expected to be within the realms of experimental verification. Acquiring the induced spin texture will require access to spin-polarized probe and low temperatures which, however, is reachable using state-of-the-art equipment.

\section{Summary and conclusions}
\label{sec-summary}
Here, we have theoretically studied local spin moment defect on a metallic Rashba surface and analyzed the effects of elastic and inelastic scattering off the local spin. We find that the response in the surface due to the elastic scattering preserves the spin characteristics of the local spin moment. Therefore, an local spin with an in-plane moment can, by elastic scattering, only induce in-plane variations in the magnetic texture of the surface electrons. However, no out-of-plane components can be generated and, hence, there is no topological change in the magnetic texture due to elastic scattering. A local spin with a finite out-of-plane component does, on the other hand, generate an out-of-plane component in the induced magnetic texture of the surface electrons. The induced spin texture tends to form helical or spiral characteristics, forming skyrmion-like surface electron waves emanating from the scattering center. Hence, a finite out-of-plane component in the local spin moment induces a magnetic texture with a different topology compared to the one of the bare Rashba surface electrons.

The situation is dramatically changes when spin-inelastic scattering is included into the picture. We show that the spin-inelastic scattering generate magnetic texture in directions orthogonal to the moments of both the surface electrons as well as the local spin. Hence, even for a local spin with only in-plane moment, inelastic scattering induces magnetic texture in the surface electrons which have finite out-of-plane components. Therefore, by turning the inelastic scattering on and off it can be used to make controlled shifts between trivial and non-trivial topology of the surface magnetic structure, something that may be potentially important both for fundamental studies of different topologies in physical systems as well as for switching and sensor applications.

By using state-of-the-art local probing techniques, e.g., spin-polarized STM, it should be possible to measure the effects discussed and predicted here.

\acknowledgements
The author thanks A. V. Balatsky, A. Bergman, J. Bouaziz, O. Eriksson, and S. Lounis for fruitful discussions. This work was supported by Vetenskapsr\aa det.

\appendix
\section{Inelastic correction to $\bfG$}
The correction $\delta\bfG(\bfr,\bfr')=\delta G_0(\bfr,\bfr')\sigma^0+\delta\bfG_1(\bfr,\bfr')\cdot\bfsigma$ to the dressed GF, including the self-energy which contains the effects from inelastic scattering, can be written as
\begin{widetext}
\begin{subequations}
\begin{align}
\delta G_0(\bfr,\bfr')=&
	\frac{1}{\pi}
	\int
		\Bigl(
			A(\bfr,\bfr_0;\dote{})G^{(0)}_0(\bfr_0,\bfr')
			+
			\bfB(\bfr,\bfr_0;\dote{})\cdot\bfG^{(0)}_1(\bfr_0,\bfr')
		\Bigr)
	d\dote{},
\\
\delta\bfG_1(\bfr,\bfr')=&
	\frac{1}{\pi}
	\int
		\Bigr(
			A(\bfr,\bfr_0;\dote{})\bfG^{(0)}_1(\bfr_0,\bfr')
			+
			\bfB(\bfr,\bfr_0;\dote{})G^{(0)}_1(\bfr_0,\bfr')
			+
			i\bfB(\bfr,\bfr_0)\times\bfG^{(0)}_1(\bfr_0,\bfr')
		\Bigr)
	d\dote{},
\end{align}
\end{subequations}
where 
\begin{subequations}
\begin{align}
A(\bfr,\bfr_0;\dote{})=&
	\calL(\dote{})
	\Bigl(
		G^{(0)}_0(\bfr,\bfr_0)\im G^{(0)}_0(\bfr_0;\dote{})
		+
		\bfG^{(0)}_1(\bfr,\bfr_0)\cdot\im\bfG^{(0)}_1(\bfr_0;\dote{})
	\Bigr)
	,
\\
\bfB(\bfr,\bfr_0;\dote{})=&
	\calL(\dote{})
	\Bigl(
		G^{(0)}_0(\bfr,\bfr_0)\im\bfG^{(0)}_1(\bfr_0;\dote{})
		+
		\bfG^{(0)}_1(\bfr,\bfr_0)\cdot\im G^{(0)}_0(\bfr_0;\dote{})
		+
		i\bfG^{(0)}_1(\bfr,\bfr_0)\times\im\bfG^{(0)}_1(\bfr_0;\dote{})
	\Bigr)
	.
\end{align}
\end{subequations}
\end{widetext}
From these expressions one easily obtains corrections to the local DOS $\delta N(\bfr,\omega)=-2\im\delta G_0(\bfr,\bfr;\omega)/\pi$ and magnetization density $\delta\bfM(\bfr,\omega)=-\im\delta\bfG_1(\bfr,\bfr;\omega)/\pi$.


\begin{thebibliography}{20}

\bibitem{heisenberg1926} W. Heisenberg, Z. Phys. {\bf 38}, 411 (1926).
\bibitem{dirac1926} P.A.M. Dirac, Proc. R. Soc. Lond. A {\bf 112}, 661 (1926).
\bibitem{heisenberg1928} W. Heisenberg, Z. Phys. {\bf 49}, 619 (1928).
\bibitem{ising1925} E. Ising, Z. Phys. {\bf 31}, 253 (1925).

\bibitem{bogdanov2001} A.N. Bogdanov, U.K. Rossler, Phys. Rev. Lett. {\bf 87}, 037203 (2001).
\bibitem{yu2010} X. Z. Yu, Y. Onose, N. Kanazawa, J. H. Park, J. H. Han, Y. Matsui, N. Nagaosa, and Y. Tokura, Nature, {\bf 465}, 901 (2010).
\bibitem{heinze2011} S. Heinze, K. von Bergmann, M. Menzel, J. Brede, A. Kubetzka, R. Wiesendanger, G. Bihlmayer, and S. Bl\"ugel, Nat. Phys. {\bf 7}, 713 (2011).
\bibitem{pereiro2014} M Pereiro, D. Yudin, J. Chico, C. Etz, O. Eriksson, and A. Bergman, Nat. Comms. {\bf 5}, 4815 (2014).

\bibitem{prassides1991} K. Prassides, J. Tomkinson, C. Christides, M. J. Rosseinsky, D. W. Murphy, R. C. Haddon, Nature {\bf 354}, 462-463 (1991).
\bibitem{christianson2008} A. D. Christianson, E. A. Goremychkin, R. Osborn, S. Rosenkranz, M. D. Lumsden, C. D. Malliakas, I. S. Todorov, H. Claus, D. Y. Chung, M. G. Kanatzidis, R. I. Bewley, and T. Guidi, Nature, {\bf 456}, 930 (2008).

\bibitem{lee2005} S. K. Lee, P. J. Eng, H. -K. Mao, Y. Meng, M. Newville, M. Y. Hu, and J. Shu, Nat. Mat., {\bf 4}, 851 (2005).
\bibitem{saiz2008} E. Garc'a Saiz, G. Gregori, D. O. Gericke, J. Vorberger, B. Barbrel, R. J. Clarke, R. R. Freeman, S. H. Glenzer, F. Y. Khattak, M. Koenig, O. L. Landen, D. Neely, P. Neumayer, M. M. Notley, A. Pelka, D. Price, M. Roth, M. Schollmeier, C. Spindloe, R. L. Weber, L.  van Woerkom, K. W\"unsch, and D. Riley, Nat. Phys., {\bf 4}, 940 (2008).

\bibitem{park2000} H. Park, J. Park, A. K. L. Lim, E. H. Anderson, A. P. Alivisatos, and P. L. McEuen, Nature, {\bf 407}, 57 (2000).
\bibitem{wang2004} W. Wang, T. Lee, I. Kretzschmar, and M. A. Reed, Nano Lett. {\bf 4}, 643 (2004).
\bibitem{rau2014} I. G. Rau, S. Baumann, S. Rusponi, F. Donati, S. Stepanow, L. Gragnaniello, J. Dreiser, C. Piamonteze, F. Nolting,
S. Gangopadhyay, O. R. Albertini, R. M. Macfarlane, C. P. Lutz, B. A. Jones, P. Gambardella, A. J. Heinrich, and H. Brune, Science, {\bf 344}, 988 (2014).

\bibitem{meier2008} F. Meier, L. Zhou, J. Wiebe, and R. Wiesendanger, Science, {\bf 320}, 82 (2008).
\bibitem{zhou2010} L. Zhou, J. Wiebe, S. Lounis, E. Vedmedenko, F. Meier, S. Bl\"ugel, P. H. Dederichs, and R. Wiesendanger, Nat. Phys. {\bf 6}, 187 (2010).

\bibitem{stipe1998} B. C. Stipe, M. A. Rezaei, and W. Ho, Science, {\bf 280}, 1732 (1998).
\bibitem{hirjibehedin2006} C. F. Hirjibehedin, C. P. Lutz, and A. J. Heinrich, Science, {\bf 312}, 1021 (2006).
\bibitem{hirjibehedin2007} C. F. Hirjibehedin, C. -Y. Lin, A. F. Otte, M. Ternes, C. P. Lutz, B. A. Jones, and A. J. Heinrich, Science, {\bf 317}, 1199 (2007).
\bibitem{otte2008} A. F. Otte, M. Ternes, K. von Bergmann, S. Loth, H. Brune, C. P. Lutz, C. F. Hirjibehedin, and A. J. Heinrich, Nat. Phys. {\bf 4}, 847 (2008).
\bibitem{otte2009} A. F. Otte, M. Ternes, S. Loth, C. P. Lutz, C. F. Hirjibehedin, and A. J. Heinrich, Phys. Rev. Lett. {\bf 103}, 107203 (2009).
\bibitem{balashov2009} T. Balashov, T. Schuh, A. F. Tak\'acs, A. Ernst, S. Ostanin, J. Henk, I. Mertig, P. Bruno, T. Miyamachi, S. Suga, and W. Wulfhekel, Phys. Rev. Lett. {\bf 102}, 257203 (2009).
\bibitem{khajetoorians2011} A. A. Khajetoorians, S. Lounis, B. Chilian, A. T. Costa, L. Zhou, D. L. Mills, J. Wiebe, and R. Wiesendanger, Phys. Rev. Lett. {\bf 106}, 037205 (2011).
\bibitem{khajetoorians2013} A. A. Khajetoorians, T. Schlenk, B. Schweflinghaus, M. dos Santos Dias, M. Steinbrecher, M. Bouhassoune, S. Lounis, J. Wiebe, and R. Wiesendanger, Phys. Rev. Lett. {\bf 111}, 157204 (2013).

\bibitem{donati2013} F. Donati, Q. Dubout, G. Aut\'es, F. Patthey, F. Calleja, P. Gambardella, O. V. Yazyev, and H. Brune, Phys. Rev. Lett. {\bf 111}, 236801 (2013).

\bibitem{heinrich2013} B. W. Heinrich, L. Braun, J. I. Pascual, and K. J. Franke, Nat. Phys. {\bf 9}, 765 (2013).

\bibitem{hasegawa1993} Y. Hasegawa and Ph. Avouris, Phys. Rev. Lett. {\bf 71}, 1071 (1993).
\bibitem{sprunger1997} P. T. Sprunger, L. Petersen, E. W. Plummer, E. L\ae gsgaard, and F. Besenbacher, Science, {\bf 275}, 1764 (1997).

\bibitem{balatsky2003} A. V. Balatsky, Ar. Abanov, and J.-X. Zhu, Phys. Rev. B, {\bf 68}, 214506 (2003).
\bibitem{fransson2007} J. Fransson and A. V. Balatsky, Phys. Rev. B, {\bf 75}, 195337 (2007).
\bibitem{she2013} J. -H. She, J. Fransson, A. R. Bishop, and A. V. Balatsky, Phys. Rev. Lett. {\bf 110}, 026802 (2013).
\bibitem{fransson2013} J. Fransson, J. -H. She, L. Pietronero, and A. V. Balatsky, Phys. Rev. B, {\bf 87}, 245404 (2013).
\bibitem{fransson2012} J. Fransson and A. V. Balatsky, Phys. Rev. B, {\bf 85}, 161401(R) (2012).
\bibitem{gawronski2010} H. Gawronski, J. Fransson, and K. Morgenstern, Nano Lett. {\bf 11}, 2720 (2011).

\bibitem{koralek2009} J. D. Koralek, C. P. Weber, J. Orenstein, B. A. Bernevig, S.-C. Zhang, S. Mack, D. D. Awschalom, Nature, {\bf 458}, 610 (2009).
\bibitem{studer2010} M. Studer, M. P. Walser, S. Baer, H. Rusterholz, S. Sch\"on, D. Schuh, W. Wegscheider, K. Ensslin, and G. Salis, Phys. Rev. B, {\bf 82}, 235320 (2010).

\bibitem{tersoff1983} J. Tersoff and D. R. Hamann, Phys. Rev. Lett. {\bf 50}, 1998 (1983).
\bibitem{wortmann2001} D. Wortmann, S. Heinze, Ph. Kurz, G. Bihlmayer, and S. Bl\"ugel, Phys. Rev. Lett. {\bf 86}, 4132 (2001).
\bibitem{fransson2010} J. Fransson, O. Eriksson, and A. V. Balatsky, Phys. Rev. B, {\bf 81}, 115454 (2010).

\bibitem{fiete2001} G. A. Fiete, J. S. Hersch, E. J. Heller, H. C. Manoharan, C. P. Lutz, and D. M. Eigler, Phys. Rev. Lett. {\bf 86}, 2392 (2001).
\bibitem{franssonNL2010} J. Fransson, H. C. Manoharan, and A. V. Balatsky, Nano Lett. {\bf 10}, 1600 (2010).
\bibitem{fransson2014} J. Fransson, A. M. Black-Schaffer, and A. V. Balatsky, Phys. Rev. B, {\bf 90}, 241409(R) (2014).

\bibitem{black-schaffer2015} A. Black-Schaffer, A. V. Balatsky, and J. Fransson, Phys. Rev. B, {\bf 91}, 201411(R) (2015).

\bibitem{lounis2012} S. Lounis, A. Bringer, and S. Bl\"ugel, Phys. Rev. Lett. {\bf 108}, 207202 (2012).


\bibitem{ast2007} C. R. Ast, G. Wittich, P. Wahl, R. Vogelgesang, D. Pacil\'{e}, M. C. Falub, L. Moreschini, M. Papagno, M. Grioni, and K. Kern, Phys. Rev. B, {\bf 75}, 201401(R) (2007).

\bibitem{TN} The topological number is defined as the integral over all space. Here, since the exact nature of the topology is not sought, it is only necessary to distinguish between vanishing and non-vanishing topological number. The numerical evaluation is therefore confined to the $10\times10$ nm area around the impurity potential.

\bibitem{hoffman2002} J. E. Hoffman, E. W. Hudson, K. M. Lang, V. Madhavan, H. Eisaki, S. Uchida, J. C. Davis, Science, {\bf 295}, 466 (2002).
\bibitem{zhu2004} J.-X. Zhu, J. Sun, Q. Si, and A.V. Balatsky, Phys. Rev. Lett. {\bf 92}, 017002 (2004).
\bibitem{strozecka2011} A. Str\'{o}\.{z}ecka, A. Eiguren, and J. I. Pascual, Phys. Rev. Lett. {\bf 107}, 186805 (2011).

\end{thebibliography}
\end{document}